\documentclass[%
 reprint,
superscriptaddress,
 amsmath,amssymb,
 aps,
]{revtex4-2}

\usepackage{graphicx}
\usepackage{dcolumn}
\usepackage{bm}
\usepackage{hyperref}

\begin{document}

\preprint{APS/123-QED}

\title{Heterogeneous node copying from hidden network structure}

\author{Max Falkenberg}
\email{max.falkenberg13@imperial.ac.uk}

\affiliation{%
 Centre for Complexity Science, Imperial College London, SW7 2AZ, U.K.
}%

\date{\today}

\begin{abstract}
Node copying is an important mechanism for network formation, yet most models assume uniform copying rules. Motivated by observations of heterogeneous triadic closure in real networks, we introduce the concept of a hidden network model -- a generative two-layer model in which an observed network evolves according to the structure of an underlying hidden layer -- and apply the framework to a model of heterogeneous copying. Framed in a social context, these two layers represent a node's inner social circle, and wider social circle, such that the model can bias copying probabilities towards, or against, a node's inner circle of friends. Comparing the case of extreme inner circle bias to an equivalent model with uniform copying, we find that heterogeneous copying suppresses the power-law degree distributions commonly seen in copying models, and results in networks with much higher clustering than even the most optimum scenario for uniform copying. Similarly large clustering values are found in real collaboration networks, lending empirical support to the mechanism. 
\end{abstract}

\maketitle

\section{Introduction}

Node copying is an important network growth mechanism \cite{newman2003structure,bhat2016densification,lambiotte2016structural,bhat2014emergence,bianconi2014triadic,davidsen2002emergence,hassan2017degree}. In social networks, copying is synonymous with triadic closure, playing an important role in the emergence of high clustering \cite{toivonen2006model,asikainen2020cumulative}. In biology, node copying encapsulates duplication and deletion, a key mechanism in the formation of protein-interaction networks \cite{farid2006evolving,pastor2003evolving,chung2003duplication,ispolatov2005duplication,bhan2002duplication}.

Despite this range of applications, most node copying models assume uniform, or homogeneous copying, i.e., that the probability of copying any given neighbour of a node is equal. The exact formulation varies widely, but examples include ``links are attached to neighbours of [node] $j$ with probability $p$'' \cite{bianconi2014triadic}, or ``one node [is duplicated]... edges emanating from the newly generated
[node] are removed with probability $\delta$'' \cite{pastor2003evolving}. Many other models use similar uniform copying rules \cite{bhat2016densification,lambiotte2016structural,krapivsky2005network,steinbock2017distribution,vazquez2003modeling,holme2002growing,vazquez2003growing,toivonen2006model,peixoto2021disentangling,battiston2016emergence,goldberg2015,davidsen2002emergence,hassan2017degree,farid2006evolving,chung2003duplication,li2013degree,bebek2006degree,steinbock2019analyticalpaths,steinbock2019analyticaldist}. 

Homogeneous copying is a sensible base assumption, often aiding a model's analytical tractability. However, especially in a social context, there is good reasons to believe that node copying may be heterogeneous. As an example, consider the social brain hypothesis, a theory which suggests that the average human has around 150 friends (Dunbar's number), encapsulating progressively smaller sub-groups of increasing social importance \cite{dunbar1998thesocial,mccarron2016calling}. In contrast, large social networks often have an average degree far exceeding Dunbar's number \cite{mcclain2017practices}, implying that most of these observed friends are only distant acquaintances. In this context, if individual A introduces individual B to one their friends, C, (i.e., B is copying A's friend C), we may reasonably expect that C is more likely to be chosen from A's inner social circle, than A's wider social circle. 

This is directly related to the principle of strong triadic closure: ``If a node A has edges to nodes B and C, then the B-C edge is especially likely
to form if A’s edges to B and C are both strong ties'' \cite{easley2010networks}. In weighted networks where tie strength can be equated to edge weight, empirical evidence for the strong triadic closure principle can be inferred by measuring the neighbourhood overlap between two nodes as a function of tie strength \cite{easley2010networks}; for example using mobile communication networks \cite{onnela2007structure}, or using face to face proximity networks \cite{scholz2014predictability}.

Unfortunately, for many networks tie strength data is unavailable or unknown. In these cases, evidence for asymmetric triadic closure may  be inferred through proxy means. For instance, in academic collaboration networks it has been shown that the ratio of triadic closure varies strongly with the number of shared collaborators between nodes \cite{kim2017over}. Although the average triadic closure ratio is small (typically $<10\%$), the ratio rapidly increases with the number of shared collaborators. However, these aggregate measures are highly coarse grained and likely only approximate real closure dynamics.

This motivates the study of simple heterogeneous copying models \cite{asikainen2020cumulative,zhou2018dynamic,bhat2014emergence,raducha2018coevolving,bianconi2014triadic}. Typically these models fall into a small number of distinct categories. In the first, heterogeneity is introduced as a node intrinsic property (e.g., node fitness) in the absence of structural considerations \cite{bianconi2014triadic}. In the second, heterogeneity is introduced via group homophily where the probability of triadic closure between nodes A and B is dependent on whether nodes A and B are in the same group or different groups (e.g., researchers from the same academic discipline, as opposed to different disciplines) \cite{asikainen2020cumulative, raducha2018coevolving}. However, intra-group copying is typically modelled uniformly. Finally, some models consider heterogeneous copying driven by the network structure around nodes A and B, without introducing node homophily \cite{bhat2014emergence}.

\citeauthor{bhat2014emergence} \cite{bhat2014emergence} define a threshold model where node A introduces node B to one of their friends C. An edge then forms between B and C if the fraction of neighbours common to B and C exceeds some threshold $F$. The model demonstrates a transition from a state where networks are almost complete for small $F$, to a state where networks are sparse but highly clustered as $F$ increases past a critical threshold. However, the model is limited in its tractability and has peculiarities such as the observation that fringe communities are almost always complete.

In this paper, our aim is to extend these ideas and introduce a more general framework for heterogeneous node copying based on the concept of hidden strong ties. To do so, we introduce the hidden network model, a framework based on multilayer networks \cite{bianconi2018multilayer} where layers have identical node structure but different edge structure. The framework lets us build models where local heterogeneity in the rules of network growth is a property of the hidden network structure and not arbitrarily encoded using node intrinsic properties or group homophily. The concept is closely related to other multilayer paradigms including the use of replica nodes to model heterogeneity \cite{cellai2016multiplex}, interdependent networks \cite{gao2012networks}, and multilayer copying \cite{battiston2016emergence}. 

In the remainder of this paper, we define and analytically study the case of extreme heterogeneous copying, the correlated copying model (CCM). The CCM is an adaptation of the uniform copying model introduced by \citeauthor{lambiotte2016structural} \cite{lambiotte2016structural}. Relaxing the extreme copying case, we numerically investigate a generalised form of the correlated copying model (GCCM) which interpolates between the UCM and CCM. 

The GCCM generates a diverse spectrum of network structures spanning both ergodic sparse and non-ergodic dense networks, with degree distributions ranging from exponential decay, through stretched-exponentials and power-laws, to extremely fat tailed distributions with anomalous fluctuations. These networks exhibit a broad clustering spectrum from sparse networks with significantly higher clustering than their uniform equivalents, to the unusual case where networks are almost complete, but with near zero clustering. We comment on a selection of real collaboration networks, which, in line with the CCM, exhibit higher clustering than can be explained by uniform copying. This suggests that heterogeneous copying may be an important explanatory mechanism for social network formation.  

\subsection{Uniform Copying Model.} 
\begin{figure}[h]
    \centering
    \includegraphics[width=0.8 \linewidth]{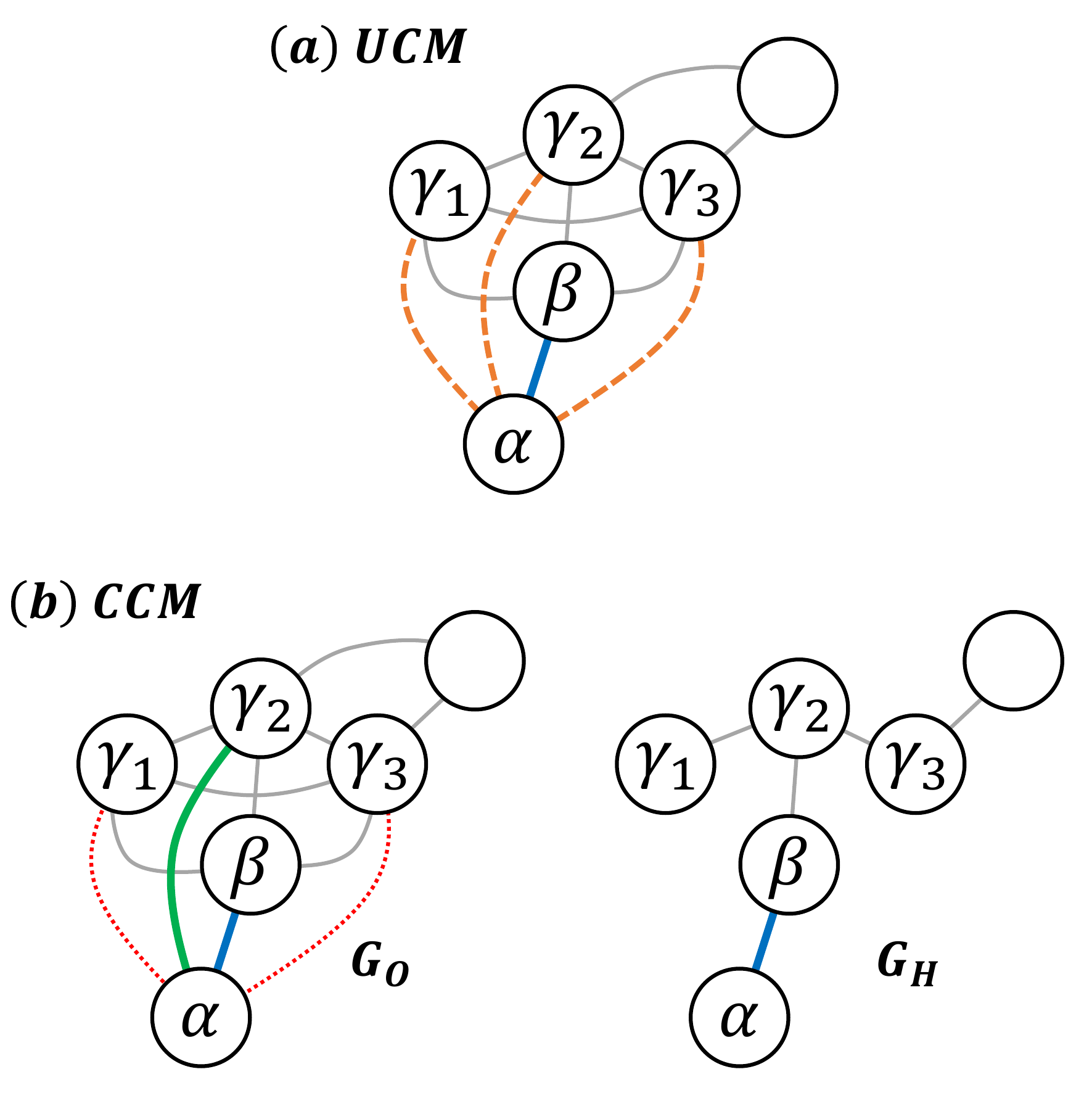}
    \caption{Two models of network formation via node copying. (a) The uniform copying model (UCM), and (b) the correlated copying model (CCM). The UCM consists of a single layer. The CCM has an observed layer, in which copying takes place, $G_O$, and a hidden layer, $G_H$. For both models, a new node $\alpha$ (the daughter) is added to the existing network (nodes connected by grey edges) and forms a random link (blue) to a target node, $\beta$ (the mother). (a) In the UCM, there is a uniform probability, $p$, of forming an edge to each of $\beta$'s neighbours ($\gamma_1,\gamma_2,\gamma_3$; orange dashed edges). (b) In the CCM, copied edges are added to the observed network, $G_O$, deterministically. If an edge exists in the hidden network, $G_H$, between node $\beta$ and node $\gamma_j$ (e.g., the $\{\beta,\gamma_2\}$ edge), then node $\alpha$ copies that edge in $G_O$ (e.g., forming the $\{\alpha,\gamma_2\}$ edge; solid green). If an edge does not exist in $G_H$ (e.g., the $\{\beta,\gamma_1\}$ and $\{\beta,\gamma_3\}$ edges), the corresponding edges are not copied to $G_O$ (red dotted lines). Copied edges are never added to $G_H$.}
    \label{fig:fig2}
\end{figure}

The uniform copying model (UCM) was introduced by \citeauthor{lambiotte2016structural} \cite{bhat2016densification,lambiotte2016structural}, see Fig.~\ref{fig:fig2}(a).
At time $t_\alpha$, a single node, $\alpha$, is added to the network,  and connects to one target node, $\beta$, which is chosen uniformly at random. The formation of an edge between the new node and the target node puts the UCM in the class of corded copying models; \citeauthor{steinbock2017distribution} \cite{steinbock2017distribution} refer to the UCM as the corded node duplication model. We label each neighbour of $\beta$ with the index $\gamma_j$ where $j \in \{1,\cdots,k^\beta\}$, and $k^\beta$ is the degree of node $\beta$. For each neighbour $\gamma_j$, the copied edge $(\alpha,\gamma_j)$ is added to the network independently with probability $p$. Following the convention of previous copying models, the nodes $\alpha$ and $\beta$ are sometimes referred to as the daughter and mother nodes respectively. The network is initialised at $t=1$ with a single node. If $p=0$, no edges are copied resulting in a random recursive tree. If $p=1$, the UCM generates a complete graph.

\section{Results \& Discussion}
\subsection{Hidden Network Models}
We define a hidden network model as the pair of single layer graphs $G = (G_O,G_H)$, comprising an observed network $G_O = (V,E_O)$ and a hidden network $G_H = (V,E_H)$, where $V$ is the set of nodes for both networks and $E_O$ and $E_H$ are the set of edges for each network. The set $V$ represent the same entities in both $G_O$ and $G_H$, with differences lying exclusively in the edge structure between nodes. The key feature of a hidden network model is that the evolution of $G_O$ is dependent on $G_H$ (or vice versa). Mathematically, this is closely related to interdependent networks \cite{danziger2014introduction}.

\subsection{Correlated Copying Model} In the correlated copying model (CCM), see Fig.~\ref{fig:fig2}(b), the observed and hidden networks are initialised with a single node at $t=1$. At $t = t_\alpha$, node $\alpha$ is added to both networks and a single target node, $\beta$, is chosen uniformly at random. We label the $k_O^\beta$ neighbours of $\beta$ in $G_O$ with the index $\gamma_j$. Then, in the observed network only, the copied edge $(\alpha,\gamma_j)$ is formed with $p_{\text{hid}}=1$ if the edge $(\beta,\gamma_j) \in E_H$, $p_{\text{obs}}=0$ otherwise. The general case with intermediate copying probabilities is discussed in section~\ref{sec:gccm}. No copied edges are added to the hidden network $G_H$. The direct edge $(\alpha,\beta)$ is added to both $G_O$ and $G_H$. The CCM therefore also falls into the class of corded node duplication models. Using the convention of referring to $\beta$ as the mother node and $\alpha$ as the daughter node, we note that the hidden network consists exclusively of first-order relations (mother-daughter), whereas edges found only in the observed network correspond to second-order relations (sister-sister, or grandmother-granddaughter).

$G_H$ evolves as a random recursive tree. Unlike the UCM, all copying in $G_O$ is deterministic, with the only probabilistic element emerging in the choice of the target node $\beta$. For comparative purposes, we define the effective copying probability in the CCM as $p_{\text{eff}} = \langle k_H^\beta/k_O^\beta\rangle$, i.e., the fraction of the observed neighbours of node $\beta$ which are copied by node $\alpha$.

Framed in a social context, we might think of $G_O$ as an observed social network where individuals have many friends, but the quality of those friendships is unknown, with most ties being weak. In contrast, underlying every social network is a hidden structure representing the inner social circle of individuals, where a node is only connected to their closest friends \cite{mccarron2016calling}. Copying in the CCM is biased to this inner circle. 

\subsubsection{Basic topological properties} The total number of edges in $G_H$ scales as $E_H(t) \sim t$, with the average degree given by $\langle k_H \rangle = 2$. Using the degree distribution of $G_H$, see below, $\langle k_H^2 \rangle = 6$. In the observed network, each time step a single edge is added by direct attachment, and one copied edge is added for each neighbour of the target node in $G_H$, $k_H^\beta$. The average change in the number of edges is therefore $\langle \Delta E_O(t) \rangle = 1 + \langle k_H^\beta \rangle = 1 + \langle k_H \rangle = 3$, such that $\langle E_O(t) \rangle \sim 3t$ and $\langle k_O \rangle = 6$. 

As an alternative, note that the observed degree of node $\alpha$ can be written as 
\begin{equation}
    (k_O)_\alpha = \sum_{\beta = 1}^{(k_H)_\alpha} (k_H)_{\alpha, \beta}
    \label{eq:hidden_obs_relation}
\end{equation}
where the index $\alpha,\beta$ labels the $(k_H)_\alpha$ unique neighbours of $\alpha$ in $G_H$. Averaging both sides of Eq.~\eqref{eq:hidden_obs_relation} over all nodes we find,
\begin{equation}
    \langle k_O \rangle =\frac{1}{t}\sum_{\alpha=1}^{t} \sum_{\beta = 1}^{(k_H)_\alpha} (k_H)_{\alpha, \beta} =\frac{1}{t}\sum_{\ell=1}^{t} n_\ell \cdot (k_H)_\ell,
    \label{eq:hidden_obs_relation_sum}
\end{equation}
where $n_\ell$ is the number of times that the degree of node $\ell$ appears in the expanded sum. For any tree graph, node $\ell$ will appear exactly once in Eq.~\eqref{eq:hidden_obs_relation_sum} for each of its $(k_H)_\ell$ neighbours. Hence, $n_\ell = (k_H)_\ell$ and $\langle k_O \rangle =\langle k_H^2 \rangle$. In supplementary note 1, Eq.~\eqref{eq:hidden_obs_relation} is used to derive $\langle k_O^2\rangle \approx 62$.

We may naively expect that the effective copying probability is $p_{\text{eff}}= \langle k_H \rangle / \langle k_O \rangle = 1/3$. However, for the CCM,  $p_{\text{eff}}= \langle k_H^\beta  / k_O^\beta \rangle \neq \langle k_H \rangle / \langle k_O \rangle$. We have not found a route to calculating this exactly, but simulations suggest $p_{\text{eff}} \approx 0.374$.

\subsubsection{Degree Distribution} The hidden network evolves as a random recursive tree which has a limiting degree distribution given by
\begin{equation}
p_H(k_H) = 2^{-k_H}, \text{ for } k_H>1.
\label{eq:hidden_degree_dist}
\end{equation}
In supplementary note 2, we show that the degree distribution for the observed network can be written as the recurrence 
\begin{equation}
p_O(k_O) =  \frac{\pi_O(k_O-1) \cdot p_O(k_O-1) + 2^{1-k_O}}{1 + \pi_O(k_O)}, \text{ for } k \geq 2,
\label{eq:masterobs2}
\end{equation}
where the final term is the probability that at time $t$ the newly added node has initial degree $k_O$ and
\begin{equation}
    \pi_O(k_O) = 1 + \langle k_H \mid k_O\rangle,
    \label{eq:degree_mapping}
\end{equation}
with $\langle k_H \mid k_O\rangle$ as the average degree of nodes in the hidden network with observed degree $k_O$.
Here, the $1$ corresponds to edges that are gained from direct attachment, whereas $\langle k_H \mid k_O\rangle$ corresponds to edges gained from copying. Although we have not found an exact expression for $\langle k_H \mid k_O\rangle$, we can make progress by considering the evolution of individual nodes.

\begin{figure}[h]
    \centering
    \includegraphics[width=\linewidth]{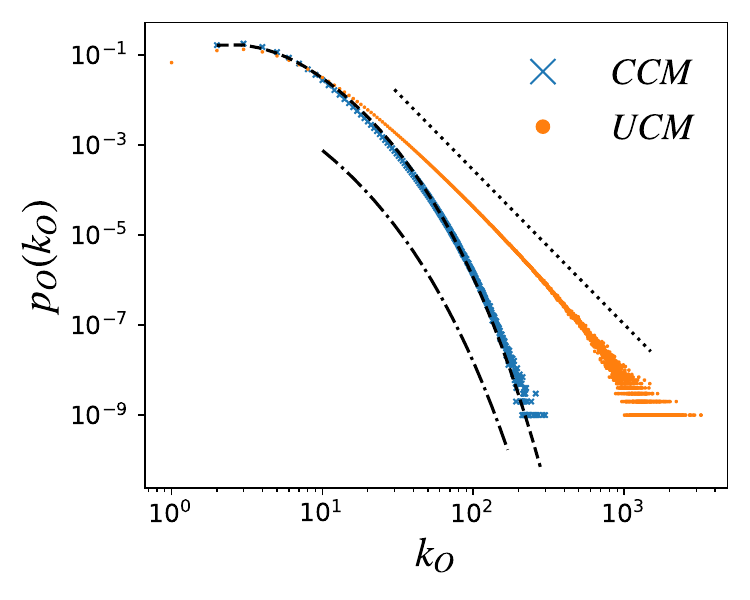}
    \caption{The degree distributions for the correlated copying model (CCM; blue crosses) and uniform copying model (UCM; orange points). Degree probability, $p_O(k_O)$, plotted as a function of the observed degree, $k_O$. UCM initialised with copying probability $p=0.374$ (equal to the CCM's effective copying probability). Networks grown to $t=10^7$, averaged over 100 networks. Error bars omitted for clarity.  Dashed line: Analytical expression for CCM in Eq.~\eqref{eq:master_ratio}. Dot-dashed: stretched exponential approximation. Dotted: power-law scaling.}
    \label{fig:degree_dist}
\end{figure}

Consider node $\alpha$ added to the network at $t_\alpha$. The initial conditions for node $\alpha$ are
\begin{subequations}
\begin{equation}
(k_H(t_\alpha))_\alpha = 1,
\end{equation}    
\begin{equation}
\langle k_O(t_\alpha)\rangle_\alpha = 1 + \langle k_H(t_\alpha - 1)\rangle_\beta,
\label{eq:obs_init}
\end{equation}
\end{subequations}
where the final term is the average hidden degree of the target node $\beta$.
In $G_H$, node $\alpha$ gains edges from direct attachment only. Hence, at $t > t_\alpha$,
\begin{equation}
\langle k_H(t)\rangle_\alpha = 1 + \sum_{j=t_\alpha}^{t-1} \frac{1}{j} = 1 + H_{t-1} - H_{t_\alpha - 1},
\label{eq:kh_scaling}
\end{equation}
where $H_n$ is the $n^{\text{th}}$ harmonic number. In $G_O$, either node $\alpha$ is targeted via direct attachment, or a copied edge is formed from the new node to node $\alpha$ via any of the $(k_H(t))_\alpha$ neighbours of node $\alpha$. Hence, 
\begin{equation}
\begin{split}
    \langle k_O(t)\rangle_\alpha &= \langle k_O(t_\alpha)\rangle_\alpha + \sum_{j=t_\alpha}^{t-1} \frac{1+\langle k_H(j)\rangle_\alpha}{j} \\
    &= \langle k_O(t_\alpha)\rangle_\alpha + \sum_{j=t_\alpha}^{t-1} \frac{2+H_{j} - H_{t_\alpha - 1} - 1/j}{j},
\label{eq:ko_scaling}
\end{split}
\end{equation}
where we have subbed in Eq.~\eqref{eq:kh_scaling} and $H_{j-1} = H_j - 1/j$. Evaluating this sum, see supplementary note 2, we find
\begin{equation}
\begin{split}
\langle k_O(t)\rangle_\alpha = \langle k_O(t_\alpha)\rangle_\alpha
+
\frac{1}{2} \Big [ (4 + H_{t-1}-H_{t_\alpha-1}&) \\  \times (H_{t-1}-H_{t_\alpha-1}) - H_{t-1}^{(2)} + H_{t_\alpha-1}^{(2)} &\Big],
\end{split}
\label{eq:obs_simple}
\end{equation}
where $H_n^{(m)}$ is the $n^{\text{th}}$ generalised Harmonic number of order $m$. For $t \rightarrow \infty$, $H_t^{(2)} \rightarrow \pi^2/6$. Hence, for large $t$ we can drop the final two terms and substitute in Eq.~\eqref{eq:kh_scaling} to give
\begin{equation}
\langle k_O(t)\rangle_\alpha \approx \langle k_O(t_\alpha)\rangle_\alpha
+
\frac{1}{2}\left( \langle k_H(t)\rangle_\alpha + 3\right)\left(\langle k_H(t)\rangle_\alpha - 1\right).
\label{eq:obs_simple3}
\end{equation}
Noting, that Eq.~\eqref{eq:obs_simple3} is a monotonically increasing function of $k_H$ for $k_H>1$, we assume that we can drop the index $\alpha$ and the time dependence giving the average observed degree of nodes with specific hidden degree as
\begin{equation}
\langle k_O \mid k_H\rangle \approx \langle \tilde{k}_O \mid k_O\rangle
+
\frac{1}{2}\left( k_H + 3\right)\left( k_H - 1\right),
\label{eq:obs_simple4}
\end{equation}
where $\langle \tilde{k}_O \mid k_O\rangle$ denotes the average initial observed degree of nodes with current degree $k_O$. Finally, we make the approximation that $\langle k_H \mid k_O\rangle \approx \langle k_O \mid k_H\rangle^{-1}$ where the exponent denotes the inverse function. This gives
\begin{equation}
\pi_O(k_O) = 1 + \langle k_H \mid k_O \rangle \approx \sqrt{2(k_O+2-\langle \tilde{k}_O \mid k_O\rangle)}.
\label{eq:mapping_gen}
\end{equation}
To proceed, let us solve the degree distribution at $k_O = 2$. Although the average initial condition $\langle \tilde{k}_O \rangle = 1 + \langle k_H \rangle = 3$, in this case $\langle \tilde{k}_O \mid 2\rangle = 2$. Therefore
\begin{equation}
p_O(2) = 
- \pi_O(2) \cdot p_O(2) + 2^{-1} = -p_O(2) \cdot \sqrt{2(2)} + 2^{-1},
\label{eq:masterobs3}
\end{equation}
giving $p_O(2) = 1/6$. Since $\langle \tilde{k}_O \mid k_O\rangle$ has an almost negligible effect on $\pi_O(k_O)$ for $k_O>2$, for simplicity we set $\langle \tilde{k}_O \mid k_O\rangle = 2$. We can now rewrite Eq.~\eqref{eq:masterobs2} as
\begin{equation}
p_O(k_O) = 
\frac{p_O(k_O - 1) \sqrt{2(k_O-1)} + 2^{1-k_O}}{1 + \sqrt{2k_O}}, \text{ for } k_O > 2. 
\label{eq:master_ratio}
\end{equation}
Although computing this recurrence shows good agreement with simulations, see Fig.~\ref{fig:degree_dist}, we have not found a closed form solution to Eq.~\eqref{eq:master_ratio}. 

As an approximation, we return to Eq.~\eqref{eq:obs_simple} and note that $H_{t-1} - H_{t_\alpha - 1} \approx \text{ln}(t/t_\alpha)$. Substituting this into Eq.~\eqref{eq:obs_simple} and dropping small terms
\begin{equation}
\langle k_O(t>t_\alpha)\rangle_\alpha \approx
2 \text{ln}(t/t_\alpha) + \frac{\text{ln}^2(t/t_\alpha)}{2},
\label{eq:ko_scaling_approx}
\end{equation}
which inverted gives
\begin{equation}
\text{ln}(t/t_\alpha) \approx -2 + \sqrt{2(k_O + 2)} \approx \sqrt{2k_O}, \text{ for } k \gg 2.
\label{eq:ko_scaling_inverted}
\end{equation}
We have dropped the expectation value and define $t_\alpha$ as the time a node was created such that its degree at time $t$ is approximately $k_O$. Exponentiating each side and taking the reciprocal,
\begin{equation}
\frac{t_\alpha}{t} \approx e^{-\sqrt{2k_O}}.
\label{eq:ko_scaling_inverted2}
\end{equation}
Finally, by substituting this approximation into the cumulative degree distribution we find
\begin{equation}
    \tilde{p}_O(k_O) = \sum_{k_O' = 2}^{k_O} p_O(k_O') \approx 1 - \frac{t_{\alpha}}{t} \approx 1 - e^{-\sqrt{2k_O}},
    \label{eq:weibull}
\end{equation}
which corresponds to a Weibull (stretched exponential) distribution, suppressing the power-law scaling observed in the UCM, see Fig.~\ref{fig:degree_dist}. 

The approximation for the cumulative degree distribution stems from the observation that, on average, nodes with $k_O' > k_O$ were added to the network at $t' < t_{\alpha}$, whereas nodes with $k_O' < k_O$ were added to the network at $t' > t_{\alpha}$. Both Eq.~\eqref{eq:ko_scaling_inverted} and Eq.~\eqref{eq:weibull} are close to the scaling expected from sub-linear preferential attachment \cite{krapivsky2000connectivity} with an exponent $1/2$.  

\subsubsection{Clique Distribution} In a simple undirected graph, a clique of size $n$ is a subgraph of $n$ nodes which is complete. A clique of size $n=2$ is an edge, whereas $n=3$ is a triangle. Here we calculate the exact scaling for the number of $n$ cliques, $Q_n(t)$, in $G_O$.

Let us first consider the case of triangles. At $t = t_\alpha$, there are two mechanisms by which a new triangle forms:
\begin{enumerate}
    \item \textbf{Direct triangles.} The new node, $\alpha$, forms a direct edge to the target node, $\beta$, and forms copied edges to each of the $k_H^\beta$ neighbours of node $\beta$, labelled with the index $\gamma_j$. The combination of the direct edge $(\alpha,\beta)$, the copied edge $(\alpha,\gamma_j)$, and the existing edge $(\beta,\gamma_j)$ creates one triangle, $(\alpha, \beta, \gamma_j)$, for each of the $k_H^\beta$ neighbours.
    \item \textbf{Induced triangles.} If node $\alpha$ forms copied edges to both node $\gamma_{j}$, and to node $\gamma_{j'}$, $j \neq j'$, the triangle $(\alpha,\gamma_{j},\gamma_{j'})$ is formed if $(\gamma_{j},\gamma_{j'}) \in E_O$. 
\end{enumerate}

\begin{figure}[h]
    \centering
    \includegraphics[width=0.78 \linewidth]{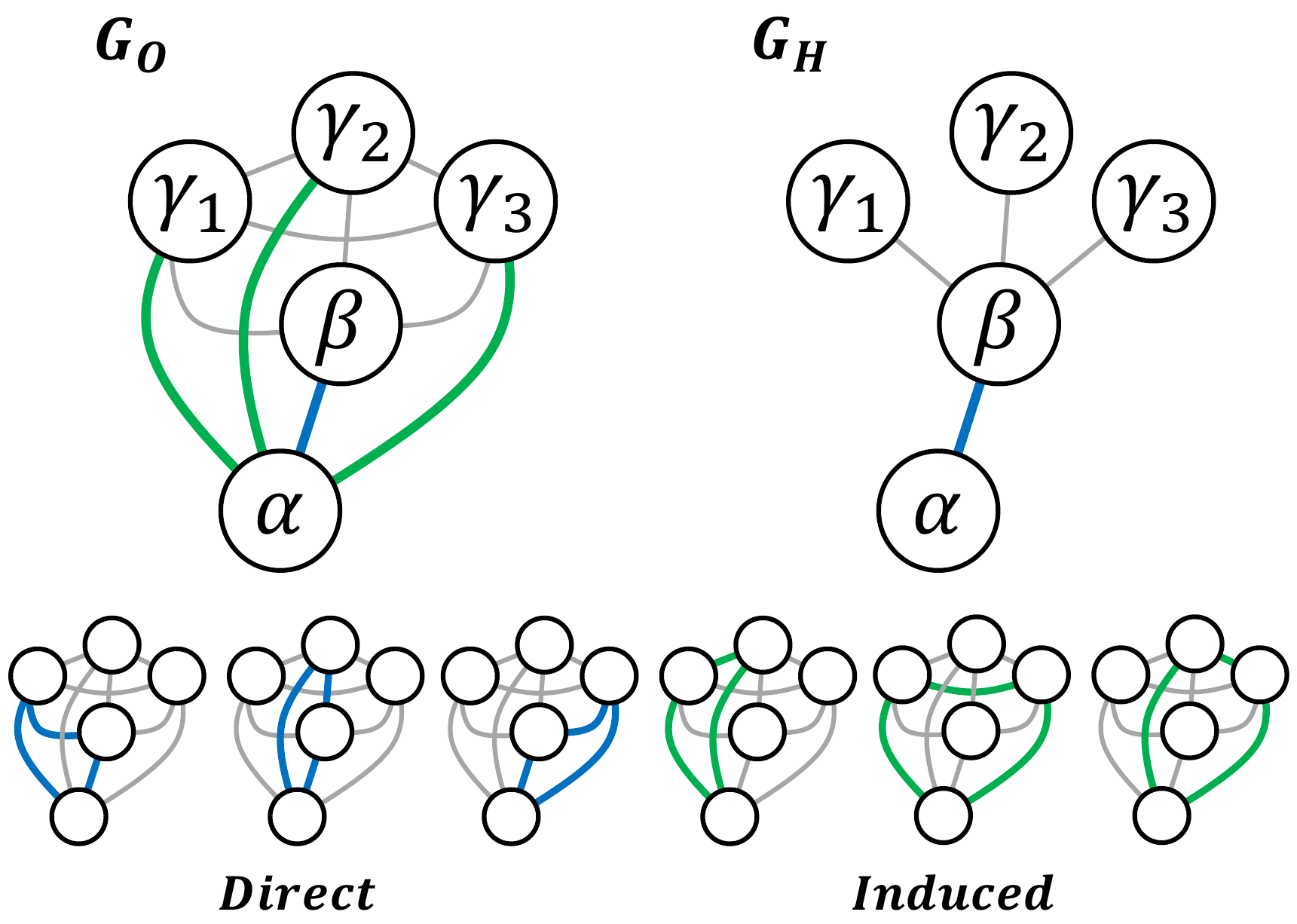}
    \caption{A schematic illustrating the number of triangles formed in a single time step of the correlated copying model. $G_O$: the observed network, $G_H$: the hidden network.
    The new node, $\alpha$, forms a direct edge (blue) to a node $\beta$ which has three existing hidden neighbours ($\gamma_1,\gamma_2,\gamma_3$). The copying process forms three new edges (green) in $G_O$. The copying process results in three new direct triangles (outlined in blue), involving the edge $\{\alpha,\beta\}$, and three new induced triangles (outlined in green), excluding the edge $\{\alpha,\beta\}$. 
    Triangles are formed in the observed network only; the hidden network remains a random tree.}
    \label{fig:clique_combinatorics}
\end{figure}

Combining these mechanisms, the change in the number of triangles can be written as
\begin{equation}
    \label{eq:triangles1}
    \Delta Q_3(t_\alpha) = \Delta Q_3^D(t_\alpha) + \Delta Q_3^I(t_\alpha),
\end{equation}
where the first and second terms on the right correspond to direct and induced triangles respectively. One new direct triangle is formed for each of the $k_H^\beta$ neighbours of node $\beta$, $\Delta Q_3^D = k_H^\beta$.
For induced triangles, the copied edge $(\alpha,\gamma_{j})$ is only formed if $(\beta,\gamma_{j}) \in E_H$. Additionally, all pairs of nodes which are next-nearest neighbours in $G_H$ must be nearest neighbours in $G_O$. Hence, the edge $(\gamma_{j},\gamma_{j'})$ must exist in the observed network if both $\gamma_{j}$ and $\gamma_{j'}$ are copied. As a result, one induced triangle is formed for each pair of copied edges $(\alpha,\gamma_{j})$ and $(\alpha,\gamma_{j'})$ such that
\begin{equation}
    \label{eq:triangle_ind}
    \Delta Q_3^I =\binom{k_H^\beta}{2}= \frac{(k_H^\beta)^2 - k_H^\beta}{2}.
\end{equation}
A visual example of the combinatorics for $k_H^\beta = 3$ is shown in Fig.~\ref{fig:clique_combinatorics}.

Extending the triangle argument to general $n$ we can write
\begin{equation}
    \label{eq:cliques1}
    \Delta Q_n(t_\alpha) = \Delta Q_n^D(t_\alpha) + \Delta Q_n^I(t_\alpha),
\end{equation}
where direct cliques are those which include the edge $(\alpha,\beta)$. For a clique of size $n$, the number of direct cliques is given by the number of ways in which $n-2$ nodes can be chosen from $k_H^\beta$ nodes,
\begin{equation}
    \label{eq:cliques_direct}
    \Delta Q_n^D(t_\alpha) = \binom{k_H^\beta}{n-2},
\end{equation}
whereas the number of induced cliques is given by the number of ways in which $n-1$ nodes can be chosen,
\begin{equation}
    \label{eq:cliques_induced}
    \Delta Q_n^I(t_\alpha) = \binom{k_H^\beta}{n-1}.
\end{equation}

As $t \rightarrow \infty$, the average change in clique number is
\begin{equation}
    \label{eq:cliques2}
    \langle \Delta Q_n(t) \rangle = \sum_{k_H = 1}^{\infty} p_H(k_H) \left [\binom{k_H}{n-2}+\binom{k_H}{n-1} \right ],
\end{equation}
where $p_H(k_H)$ is the probability that the randomly chosen target node $k_H^\beta = k_H$. To avoid ill-defined binomials, we rewrite Eq.~\eqref{eq:cliques2} as
\begin{equation}
\langle \Delta Q_n(t) \rangle = p_H(n-2) + \sum_{k_H = n-1}^{\infty} p_H(k_H) \cdot \binom{k_H+1}{n-1},
\label{eq:cliques3}
\end{equation}
where we have combined the two terms into a single binomial. After subbing in $p_H(k_H)$ and solving the sum,
\begin{equation}
    \langle \Delta Q_n(t) \rangle = 2^{2-n} + \sum_{k_H = n-1}^{\infty} 2^{-k_H} \cdot \binom{k_H+1}{n-1} = 4.
\end{equation}
Consequently, for large $t$ we find the curious result that the number of $n$ cliques scales as
\begin{equation}
    Q_n(t) \sim 4t, \text{ for } n>2,
    \label{eq:clique_scaling}
\end{equation}
independent of the clique size. In practice this result only applies for $t \rightarrow \infty$. To see this, note that the largest clique in $G_O$ at time $t$ is always directly related to the largest degree node in $G_H$,
\begin{equation}
    \text{Max}(n,t) = \text{Max}(k_H,t) + 1,
\end{equation}
with the largest hidden degree at time $t$ scaling as approximately 
\begin{equation}
    \text{Max}(k_H,t) \sim \text{ln(t)}.
\end{equation}
We can invert this and ask how large the network is if we observe that the largest observed clique is $n$. This gives
\begin{equation}
    t_n \sim e^{\text{n}}.
\end{equation}
Hence, the scaling relation in Eq.~\eqref{eq:clique_scaling}, is only valid for cliques of size $n$ when $t \gg t_n$. In supplementary note 3, we plot the number of cliques in simulations of the CCM as a function of $t$. For small clique sizes, the scaling in Eq.~\eqref{eq:clique_scaling} is clearly apparent early in the evolution of the CCM. However, for moderate and large cliques, the standard deviation in the number of cliques is significantly larger than the average number of cliques, obscuring a clear trend. 

\subsubsection{Clustering} Transitivity is a global clustering measure defined as 
\begin{equation}
    \tau_{G_O} = 3 \times \frac{\#(\text{triangles in }G_O)}{\#(\text{twigs in }G_O)},
    \label{eq:transitivity_def}
\end{equation}
where a twig is any three nodes connected by two edges. The number of twigs is equivalent to the number of star graphs of size $2$, $S_2$, where a star graph of size $n$ is a subgraph with $1$ central node and $n$ connected neighbours. The number of subgraphs of size $2$ is related to the degree distribution by
\begin{equation}
    S_2(t) = t\sum_{k_O\geq 2} \binom{k_O}{2} \cdot p_O(k_O)  = t \cdot \frac{\langle k_O^2\rangle-\langle k_O \rangle}{2} ,
    \label{eq:star_graph2}
\end{equation}
where we have used the property that $p_O(k<2) = 0$. Recalling that $\langle k_O \rangle = 6$ and $\langle k_O^2 \rangle \approx 62$, the number of twigs scales as $S_2 \sim 28t$, such that
\begin{equation}
    \tau_{G_O} = \frac{3 Q_3}{S_2} \sim \frac{3 \cdot 4t}{28t} = \frac{3}{7}.
    \label{eq:ccm_trans}
\end{equation}

The observed network can be recovered from the hidden network by converting every wedge in $G_H$ into a triangle. This can be thought of as complete triadic closure where every possible triangle which can be closed, from the addition of a single edge to the hidden network, is closed. This implies that the CCM has the largest possible transitivity from a single iteration of triadic closure on a random recursive tree.

The local clustering coefficient, $cc(\alpha)$, is defined as the number of edges between the $(k_O)_\alpha$ neighbours of $\alpha$, normalised by the the number of edges in a complete subgraph of size $(k_O)_\alpha$. For the CCM, 
\begin{equation}
    cc(\alpha) = \frac{\binom{(k_H)_\alpha}{2} + \sum_{\beta = 1}^{(k_H)_\alpha}\binom{(k_H)_{\alpha,\beta}}{2}}{\binom{(k_O)_\alpha}{2}}, 
    \label{eq:cc_local}
\end{equation}
where the first term corresponds to the complete subgraph of the $(k_H)_\alpha$ neighbours of $\alpha$ in $G_H$, and the sum contributes the edges from one complete subgraph formed by node ${\alpha,\beta}$ and its $(k_H)_{\alpha,\beta} - 1$ neighbours, excluding $\alpha$. The global clustering coefficient, $CC(G_O)$, is defined as the average of Eq.~\eqref{eq:cc_local} over all nodes in the network.
In simulations, $CC(G_O) \approx 0.771$ for large $t$.

\subsubsection{Path Lengths} \citeauthor{steinbock2017distribution} \cite{steinbock2017distribution} calculate the distribution of shortest path lengths for the UCM (referred to in their paper as the corded node duplication model). Specifically, the authors calculate the probability that two randomly chosen nodes, $i$ and $j$, will be separated by a shortest path of length $\ell$, denoted as $\mathcal{P}(L=\ell; t)$, at time $t$. 

The UCM with $p=0$ corresponds to a random recursive tree and is therefore equivalent to the $G_H$. Hence, for the hidden network, we can lift the path length distribution, $\mathcal{P}_H(L_H=\ell; t)$, and the mean shortest path, $\langle L_H(t)\rangle$, from \citeauthor{steinbock2017distribution} \cite{steinbock2017distribution}. We can then exploit a convenient mapping to calculate the distribution of shortest path lengths in $G_O$ from $G_H$. 

Consider two randomly chosen nodes $i$ and $j$. In $G_H$, there is a unique path (due to its tree structure) from $i$ to $j$ of length $(\ell_{H})_{ij}$. In $G_O$, the enforced triadic closure process means that for every two steps on the path from $i$ to $j$ in $G_H$, an observed edge exists in $G_O$ which acts as a shortcut, reducing the path length by one. Hence, if the path length $(\ell_{H})_{ij}$ is even, the path length in $G_O$ is given by $\ell_{O} = \ell_H / 2$; if the path length is odd $\ell_O = (\ell_H + 1) / 2$. Using this mapping, we can write
\begin{equation}
\begin{split}
\mathcal{P}_O(L_O=\ell; t) = \mathcal{P}_H(L_H=2\ell; t) + \mathcal{P}_H(L_H=2\ell - 1; t), \\
\text{for } \ell \geq 1.
\end{split}
\label{eq:path_length_distribution}
\end{equation}
If we assume that, for large $t$, there are an approximately equal number of odd and even shortest paths in $G_H$, the average shortest path length in $G_O$ is
\begin{equation}
    \langle L_O(t) \rangle = \frac{\langle L_H(t) \rangle}{2} + \frac{1}{4},
    \label{eq:obs_paths_average}
\end{equation}
where the $1/4$ term accounts for the discrepancy in the mapping for odd and even paths. 

From \citeauthor{steinbock2017distribution} \cite{steinbock2017distribution}, we note that the mean shortest path length for $G_H$ scales as
\begin{equation}
    \langle L_H(t) \rangle \sim 2 \cdot \text{ln}(t),
    \label{eq:hidden_paths}
\end{equation}
which indicates that the hidden network exhibits the small-world property \cite{watts1998collective}. We have omitted constants which are negligible at large $t$. Hence, applying the mapping in Eq.~\eqref{eq:obs_paths_average} and omitting the $1/4$ term for simplicity, the mean shortest path length for $G_O$ is given by
\begin{equation}
    \langle L_O(t) \rangle \sim  \text{ln}(t),
    \label{eq:obs_paths}
\end{equation}
indicating that the observed network also exhibits the small-world phenomenon. This mapping is confirmed by simulations.

For interest, we note that for $0<p<1$, the shortest paths for the UCM are in general not unique; there may be multiple paths between nodes $i$ and $j$ which are equally short. Unusually for a non-tree network, all shortest paths are unique in the CCM.

\subsection{General Correlated Copying Model \label{sec:gccm}} The general correlated copying model (GCCM), is defined analogously to the CCM, starting with observed and hidden networks initialised at $t=1$. Like the UCM and CCM, the GCCM is a corded node duplication model. For practical reasons, we initialise the graph with three nodes which form a complete graph in $G_O$, and a wedge in $G_H$. This ensures that the initial graph contains some edges found in $G_H$, and some edges found only in $G_O$.

At $t = t_\alpha$, node $\alpha$ is added to both networks and a single target node, $\beta$, is chosen uniformly at random. We label the $k_O^\beta$ neighbours of $\beta$ in $G_O$ with the index $\gamma_j$. In the observed network, the copied edge $(\alpha,\gamma_j)$ is formed with probability $p_{\text{hid}}$ if the edge $(\beta,\gamma_j) \in E_H$ (inner circle copying), and probability $p_{\text{obs}}$ otherwise (outer circle copying). The direct edge $(\alpha,\beta)$ is added to both $G_O$ and $G_H$. 

The GCCM encapsulates a wide spectrum of heterogeneous copying. Setting $p_{\text{hid}} = 1$ and $p_{\text{obs}}=0$ reduces the GCCM to the CCM, whereas setting $p_{\text{hid}} = p_{\text{obs}} = p$ reduces the GCCM to the UCM. We have discussed the social motivation for the case where $p_{\text{hid}} > p_{\text{obs}}$, representing a copying bias towards the inner social circle of a node. However, the GCCM can also be tuned to the reverse case where $p_{\text{hid}} < p_{\text{obs}}$, resulting in a bias against inner circle nodes. We are not aware of a clear physical motivation for this latter case. However, the structural diversity of these anti-correlated networks warrants their discussion here. 

\begin{figure}[h]
    \centering
    \includegraphics[width=\linewidth]{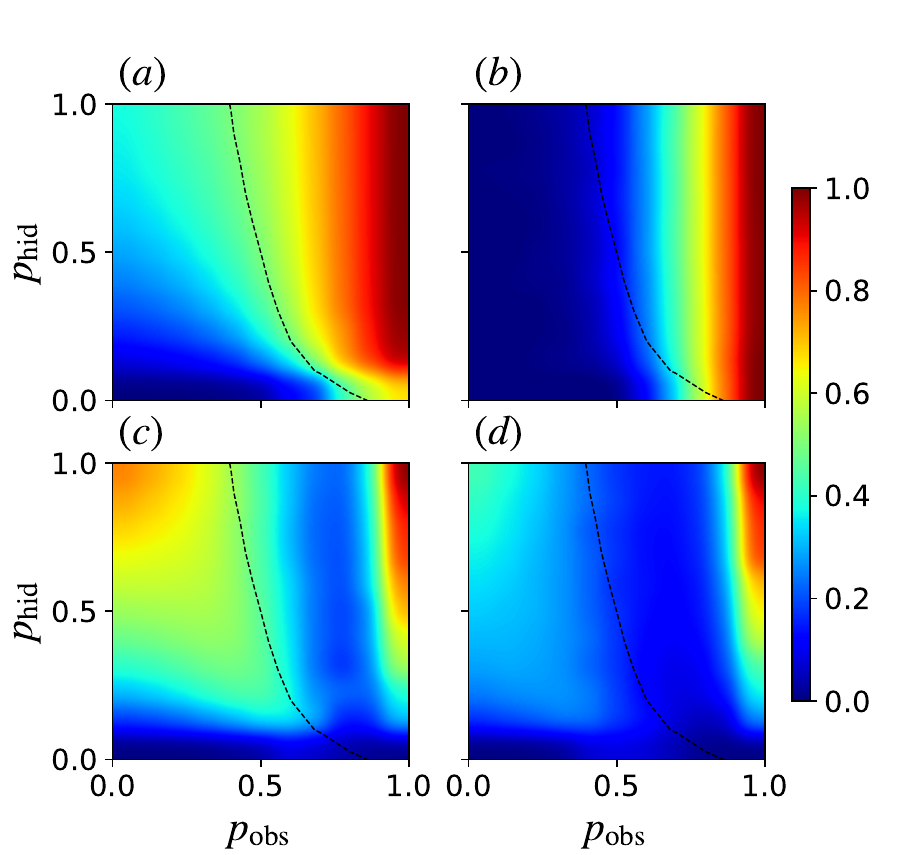}
    \caption{Properties of the observed network, $G_O$, in the general correlated copying model (GCCM). Numerical results for $10^4$ nodes as a function of the hidden copying probability, $p_{\text{hid}}$, and the outer copying probability, $p_{\text{obs}}$. (a) The effective copying probability. (b) The densification exponent. (c) The average local clustering coefficient, $CC(G_O)$. (d) The transitivity, $\tau_{G_O}$. Black dashed contour: effective copying probability of $0.5$ at $t=10^4$, calculated numerically. Values have been smoothed for clarity.}
    \label{fig:gccm_r0}
\end{figure}

Figure~\ref{fig:gccm_r0} shows numerical results for (a) the effective copying probability, (b) the densification exponent, (c) the average local clustering coefficient, and (d) the transitivity, for the GCCM with $10^4$ nodes. 

The effective copying probability corresponds to the fraction of target node neighbours which appear to be copied in the observed network. Formally, we can write the average effective copying probability at time $t$ as
\begin{equation}
    p_{\text{eff}}(t) = \Biggl\langle\frac{p_{\text{hid}} k_H^\beta + p_{\text{obs}} (k_O^\beta - k_H^\beta)}{k_O^\beta}\Biggr\rangle,
\end{equation}
where $\beta$ is the index of the target node at time $t$, the first term represents edges copied from node $\beta$'s inner circle, and the second term represents edges copied from the outer circle. 

The dashed contour in Fig.~\ref{fig:gccm_r0}(b) corresponds to an effective copying probability of $0.5$, calculated numerically by averaging over the preceding $10^4$ time steps. We note that $p_{\text{eff}} = 0$ if $p_{\text{hid}} = p_{\text{obs}} = 0$ (random tree), $p_{\text{eff}} = 1$ if $p_{\text{hid}} = p_{\text{obs}} = 1$ (complete graph), and $p_{\text{eff}} = p$ if $p_{\text{hid}} = p_{\text{obs}} = p$ (UCM). In general, the rise in $p_{\text{eff}}$ is faster with increasing $p_{\text{obs}}$ than increasing $p_{\text{hid}}$, although for $p_{\text{hid}} =0$ we find very small $p_{\text{eff}}$, even for large $p_{\text{obs}}$. However, this observation is somewhat deceptive since, if the GCCM is in the dense regime and $p_{\text{hid}} \neq p_{\text{obs}}$, $p_{\text{eff}}$ is not stationary. Calculated over longer time frames, we note that the effective copying probability appears to slowly converge to the outer copying probability, $p_{\text{eff}}\rightarrow p_{\text{obs}}$, since for $t\rightarrow \infty$, the ratio of the number of edges in the hidden network to the number of edges in the observed network tends to zero. This suggests that the dashed $p_{\text{eff}}=0.5$ contour will converge to the $p_{\text{obs}}=0.5$ line as $t \rightarrow \infty$.

We test whether the GCCM is in the sparse or dense regime explicitly by tracking the growth in the number of edges in the observed network. Let us define the densification exponent, $\delta$, $0 \leq \delta \leq 1$, using $E_O(t) \propto t^{1+\delta}$, which relates the number of edges in the observed network to the number of nodes $t$. If $\delta \approx 0$, the GCCM is sparse. If $\delta = 1$, the GCCM grows as a complete graph. For intermediate values, the GCCM undergoes densification. For the UCM, the transition from the sparse to dense regime is known to take place at $p=0.5$ \cite{lambiotte2016structural}. We have not analytically calculated the transition for the GCCM, but may intuitively expect the transition at $p_{\text{obs}} = 0.5$ since the hidden network is a random tree. This seems to be supported by the numerical values of $\delta$ in Fig.~\ref{fig:gccm_r0}(b), although the transition from zero to non-zero $\delta$ is shifted to slightly larger $p_{\text{obs}}$ for $p_{\text{hid}} = 0$, and to smaller $p_{\text{obs}}$ for $p_{\text{hid}} = 1$; this shift is likely to disappear as $t\rightarrow \infty$.

Figures~\ref{fig:gccm_r0}(c) and (d) show the average local clustering coefficient, $CC(G_O)$, and transitivity (global clustering), $\tau_{G_O}$, for the GCCM. Patterns are similar between the two figures, although local clustering generally exceeds global clustering in the sparse regime. For the UCM it is known that, in the dense regime, $\tau_{G_O}$ slowly converges to zero as $t\rightarrow \infty$, unless $p=1$ \cite{bhat2016densification}. In contrast, the local clustering appears to remain non-zero. 

As expected, clustering is minimised at $p_{\text{hid}} = p_{\text{obs}} = 0$ (random tree) and maximised for a complete graph,  $p_{\text{hid}} = p_{\text{obs}} = 1$. However, in the sparse regime we find that the maximum clustering is found at $p_{\text{hid}} = 1$,  $p_{\text{obs}} = 0$ which corresponds to the CCM. \citeauthor{bhat2016densification} \cite{bhat2016densification} note that local and global clustering for the UCM is not a monotonically increasing function of the copying probability $p$, with a local maxima in the sparse regime at non-zero $p$. This bimodal clustering is also present in the GCCM. In the anti-correlated regime where $p_{\text{hid}} \approx 0$, we find near zero clustering values. In particular if $p_{\text{hid}} = 0$ and $p_{\text{obs}} = 1$, we observe the unusual property that $\delta \approx 1$, such that the network scales as (but is not) a complete graph, yet both the local and global clustering are approximately zero. 

\begin{figure*}[t]
    \centering
    \includegraphics[width=\linewidth]{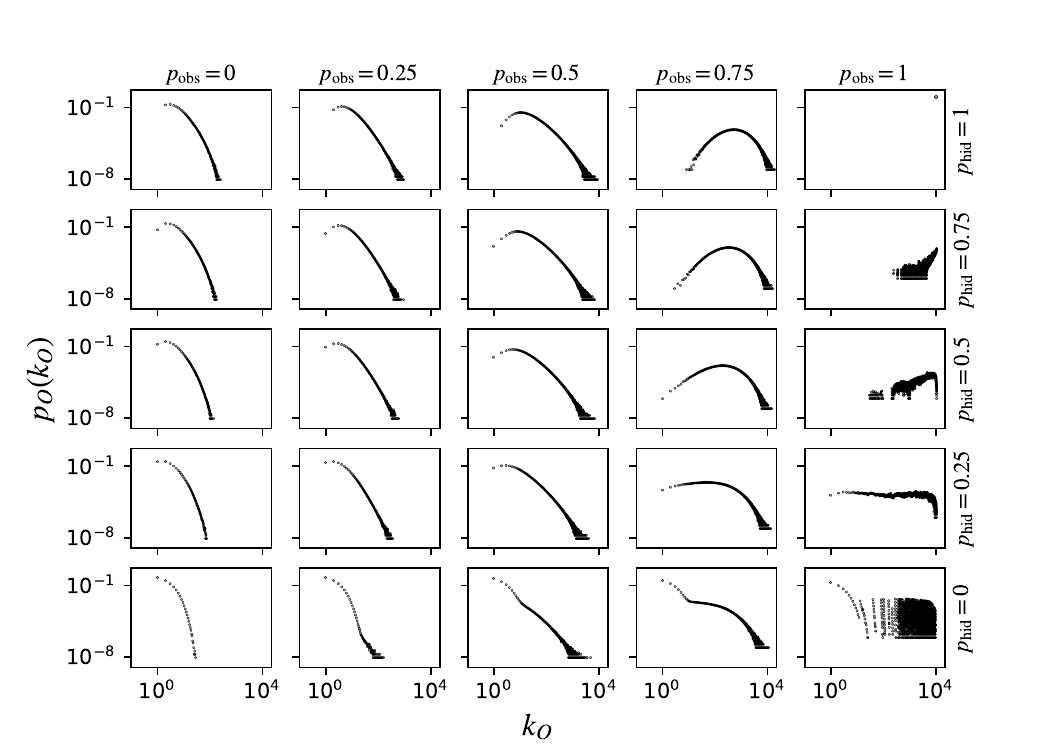}
    \caption{The observed degree distributions for the general correlated copying model (GCCM). The degree probability, $p_O(k_O)$, is plotted as a function of the observed degree, $k_O$, for various values of the outer copying probability, $p_{\text{obs}}$ (left to right), and the hidden copying probability, $p_{\text{hid}}$ (bottom to top). For $p_{\text{obs}} \in \{0,0.25,0.5\}$, each network contains $10^6$ nodes. For $p_{\text{obs}} = 0.75$, each network contains $10^5$ nodes. For $p_{\text{obs}}=1$, each network contains $10^4$ nodes. Distributions are averaged over 100 instances. In the dense regime ($p_{\text{obs}}>0.5$), network growth is non-ergodic leading to anomalous scaling and noisy degree distributions. The distribution at $p_{\text{hid}}=p_{\text{obs}}=0$ corresponds to a random recursive tree, see Eq.~\ref{eq:hidden_degree_dist} (exponential decay). The distribution at $p_{\text{hid}}=p_{\text{obs}}=1$ corresponds to a complete graph. The distribution at $p_{\text{hid}}=1$, $p_{\text{obs}}=0$ corresponds to the CCM, see Fig.~\ref{fig:degree_dist}. If $p_{\text{hid}}=p_{\text{obs}}$, the GCCM is equivalent to the uniform copying model.}
    \label{fig:degree_dist_all}
\end{figure*}

Extracting the degree distributions for the GCCM for various $p_{\text{hid}}$ and $p_{\text{obs}}$ shows similarly diverse behaviour, see Fig.~\ref{fig:degree_dist_all}. Each distribution is averaged over 100 instances, but points are left deliberately unbinned to illustrate the significant fluctuations observed in the dense regime. For $p_{\text{hid}} = p_{\text{obs}} = 0$ (bottom left) the GCCM reduces to a random recursive tree, see Eq.~\eqref{eq:hidden_degree_dist}. The CCM case with $p_{\text{hid}} = 1$, $p_{\text{obs}} = 0$ (top left) follows Eq.~\eqref{eq:master_ratio}, where the tail can be approximated as a stretched exponential. This distribution is also shown in Fig.~\ref{fig:degree_dist}. Along the diagonal where $p_{\text{hid}} = p_{\text{obs}}$ (UCM), the degree distribution has a power-law tail in the sparse regime, and exhibits anomalous scaling in the dense regime ($p\geq 0.5$). For $p_{\text{hid}} = p_{\text{obs}} = 1$, the GCCM reduces to a complete graph and all nodes have degree $t-1$.

For $p_{\text{obs}} = 0$, the power-law scaling observed in the UCM is completely suppressed, with a gradual transition from exponential decay to a stretched exponential tail as $p_{\text{hid}}$ is increased from $0$ to $1$. In the sparse regime with $p_{\text{obs}} \neq 0$, all degree distributions appear fat tailed with only small deviations from the power-laws observed for the UCM. However, unusual scaling is observed for $p_{\text{hid}} = 0$, $p_{\text{obs}} \neq 0$, where the distributions exhibit initial exponential decay at small $k_O$, attributable to the hidden network, before a second fat-tailed regime starting at intermediate $k_O$. 

In the dense regime, all degree distributions exhibit anomalous scaling, such that individual instances are not self-averaging. For $p_{\text{obs}} = 0.75$, the tail of the degree distributions is largely consistent across all $p_{\text{hid}}$. However, the probability of finding nodes with small degree is large for $p_{\text{hid}} = 0$, and is gradually suppressed as $p_{\text{hid}} \rightarrow 1$. These effects are most pronounced for $p_{\text{obs}} = 1$ where the modal degree is $1$ for $p_{\text{hid}}=0$, and $t-1$ for $p_{\text{hid}}=1$, with a gradual transition in between. Throughout this transition the degree distribution appears almost uniform at $p_{\text{hid}} = 0.25$, where the probability of finding nodes with any given degree is approximately constant up until the large $k_O$ limit. However, this effect is only observed when averaging over many instances, with a much smaller degree range observed in individual networks. 

It is possible to extend the GCCM further by adding copied edges from $G_O$ to the hidden network, $G_H$, with probability $q$.
Results are shown in supplementary note 4 for $q>0$ where clustering is enhanced if $p_{\text{hid}} > p_{\text{obs}}$ and suppressed if $p_{\text{hid}} < p_{\text{obs}}$, relative to the UCM. In the limiting case of $q=1$, the GCCM is independent of $p_{\text{obs}}$ and equivalent to the UCM with $p=p_{\text{hid}}$. The $p_{\text{hid}} =p_{\text{obs}}$ line (UCM) is invariant under changes in $q$. One potential application of the $q\neq 0$ case is for generating random simplicial complexes \cite{battiston2020networks} by combining the hidden and observed networks into a single structure. Such a construction may be interesting since it explicitly distinguishes between cliques of strong ties, where all nodes are within each other's inner circle, and cliques of weak ties, see supplementary note 5.

\subsection{Comparing Copying Models}

We have introduced a simple model of heterogeneous node copying, motivated by arguments that triadic closure may not be structurally homogeneous in real networks. 

Comparing the CCM, for which we have analytical results, to the UCM with the equivalent effective copying probability ($p = p_{\text{eff}} = 0.374$) we find significant differences in network structure. Both the average local clustering coefficient, $CC(G_O)$, and the transitivity, $\tau_{G_O}$, are significantly larger in the CCM than the UCM. The CCM suppresses the power-law tail observed in the UCM for the sparse regime, and consequently, the degree variance observed in the CCM is smaller than for the UCM. CCM: $\sigma^2(k_O) \approx 26$; UCM: $\sigma^2(k_O) \approx 192$. The CCM also has the unusual property, not found in the UCM, that the growth in the number of cliques of size $n$ scales independently of $n$ as $t\rightarrow \infty$. For both the UCM and CCM, the mean shortest path lengths scale as $\sim \text{ln}(t)$ indicative of the small-world property.

The above comparison uses a single effective copying probability, but key differences are robust for variable $p$ in the sparse regime. Specifically, the UCM degree distribution always exhibits a power-law tail, and the largest measured clustering coefficients fall below the values seen for the CCM, see Tab.~\ref{tab:ccm_vs_ucm}. Relaxing the CCM to the GCCM, we note that for large $p_{\text{hid}}$ and small $p_{\text{obs}}$, the measured clustering values regularly exceed those observed in the UCM, with the UCM only reaching similar values far into the dense regime. Given the continuing debate about the ubiquity of power-laws in real networks \cite{broido2019scale}, the observation that power-laws are suppressed in the GCCM as soon as the UCM symmetry is broken supports the view that power-law network scaling is an idealised case which in practice is rarely observed for real networks. 
\begin{table}[h]
\begin{tabular}{r|l|c|c|c|}
\cline{2-5}
\multicolumn{1}{l|}{}               & $\bm{p_{\text{eff}}}$ & $\bm{CC(G_O)}$ & $\bm{\tau_{G_O}}$ & \textbf{Degree Dist.}      \\ \hline
\multicolumn{1}{|r|}{CCM}           & 0.37            & 0.77        & 0.43         & Str. Exp. \\ \hline
\multicolumn{1}{|r|}{UCM (Max $\bm{CC(G_O)}$)}  & 0.38            & 0.52        & 0.20         & Power-law        \\ \hline
\multicolumn{1}{|r|}{UCM (Max $\bm{\tau_{G_O}}$)} & 0.22            & 0.40        & 0.28         & Power-law        \\ \hline
\end{tabular}
\caption{A comparison between the correlated copying model (CCM) and the uniform copying model (UCM). UCM is simulated twice, once with an effective copying probability, $p_{\text{eff}}$, that results in the network with the highest average local clustering coefficient (in the sparse regime), $CC(G_O)$, and once with the effective copying probability that gives the largest transitivity, $\tau_{G_O}$. Values averages over 50 simulations where each network contains $10^5$ nodes. Standard deviations are negligible. All values for the observed network, $G_O$.}
\label{tab:ccm_vs_ucm}
\end{table}

Whether such extreme bias is plausible in real networks is uncertain. However, observations in academic collaboration networks suggest that extreme bias may be possible \cite{kim2017over}. For instance,  \citeauthor{kim2017over} \cite{kim2017over} show that the ratio of triadic closure between two nodes is approximately zero if the number of shared collaborators is zero,  rises rapidly as the number of shared collaborators increases, and plateaus at a ratio of one.

A second clue towards heterogeneous copying is the observation of very large clustering values in real networks. A selection of these networks and their clustering coefficients is shown in Tab.~\ref{tab:networks}. Stressing that both the UCM and GCCM are toy models of node copying, the networks in Tab.~\ref{tab:networks} exhibit average local and/ or global clustering far exceeding even the most optimistic values for the UCM. In contrast, the listed clustering values are relatively similar to what may plausibly emerge from heterogeneous copying, although even the clustering observed for the extreme CCM case falls below some of the values shown in Tab.~\ref{tab:networks}. Future work should go beyond this qualitative analysis and should attempt to measure the degree to which copying symmetry is broken for real networks where these mechanisms are relevant.
\begin{table}[h]
\begin{tabular}{r|c|c|c|c|}
\cline{2-5}
\multicolumn{1}{l|}{}                          & \textbf{Nodes} & \textbf{Edges} & $\bm{CC(G_O)}$ & $\bm{\tau_{G_O}}$ \\ \hline
\multicolumn{1}{|r|}{arXiv Astro coauthors}    & 18.8K          & 198.1K         & 0.63        & 0.32         \\ \hline
\multicolumn{1}{|r|}{arXiv GR coauthors}       & 5.2K           & 14.5K          & 0.53        & 0.63         \\ \hline
\multicolumn{1}{|r|}{arXiv CM coauthors}       & 23.1K          & 93.4K          & 0.63        & 0.26         \\ \hline
\multicolumn{1}{|r|}{arXiv HEP coauthors}      & 22.9K          & 2.7M           & 0.81        & 0.31         \\ \hline
\multicolumn{1}{|r|}{DBLP coauthors}           & 540.5K         & 15.2M          & 0.80        & 0.65         \\ \hline
\multicolumn{1}{|r|}{NetSci coauthors}         & 379            & 914            & 0.74        & 0.43         \\ \hline
\multicolumn{1}{|r|}{Hollywood collaborations} & 1.1M           & 56.3M          & 0.77        & 0.31         \\ \hline
\multicolumn{1}{|r|}{DNC Email corecipients}    & 906            & 12.1K          & 0.61        & 0.56         \\ \hline
\end{tabular}
\caption{A selection of sparse undirected networks which may plausibly grow via a copying mechanism. These networks exhibit larger average local clustering coefficients, $CC(G_O)$, and transitivity, $\tau_{G_O}$, than one may expect if these networks were to grow via a uniform copying mechanism. Network source data from \citeauthor{networksrepository2015} \cite{networksrepository2015}.}
\label{tab:networks}
\end{table}

\subsection{Discussion}
The UCM, CCM and GCCM are all examples of  corded copying models where an edge forms between a newly added node and the target node which is duplicated. This is in contrast to uncorded duplication models where a new node is formed by copying an existing target node and its neighbours, but an edge is not formed between the new node and the duplicated target node. Corded models are more common in the context of social phenomena and triadic closure, whereas uncorded models are typically more relevant to duplication--divergence processes in protein interaction networks. In the current work we have focused exclusively on corded models; considering heterogeneous copying in uncorded models \cite{bhan2002duplication,ispolatov2005duplication,bebek2006degree,li2013degree} would be an appropriate future extension. Heterogeneous copying could also be studied by extending directed models  \cite{steinbock2019analyticaldist,steinbock2019analyticalpaths}.

The GCCM and CCM are examples of hidden network models. From a mathematical standpoint, hidden network models can be thought of as a variant of interdependent networks where nodes in one layer have dependencies of nodes in another layer \cite{gao2012networks,danziger2014introduction}. However, at a conceptual level, hidden network models puts an emphasis on how the evolution of network structure can depend on asymmetries not observed in our data.

In this paper we have focused on copying in social networks, but the ideas naturally extend to other contexts. In economics, our framework may be applied to shareholder networks \cite{bardoscia2021physics}, where nodes are connected if they both own a common asset. Here, the hidden network represents the full set of co-owned assets, whereas the observed network includes publicly disclosed assets. Similarly, the idea can be applied to co-bidding networks in public procurement, where an edge indicates that two companies both placed bids on the same contract. In many jurisdictions, only winning bids (of which there may be multiple) are publicly revealed. Therefore, the observed network may represent the network of winning bids, whereas the hidden network includes all bids. Hidden network models may be a valuable representation in these cases if there are structural reasons for why some data is observed and some data is hidden. For instance, fraudulent behaviour in public procurement has been associated with anomalous structural features in the co-bidding network \cite{wachs2019network}.

Other examples may be found in ecology, where multilayer networks have been used to represent different interactions between a common set of species \cite{kefi2015network,pilosof2017multilayer}. \citeauthor{kefi2015network} \cite{kefi2015network} find that the structure of interactions in one layer has significant cross-dependencies to the structure of other layers. This mirrors how interlayer dependencies in the CCM are used to break symmetries in the evolution of the observed network.
Finally, hidden networks may find general relevance to other fields where interdependent networks have been influential. This may include studies on energy demand management for power grids \cite{iacopini2020multilayer}, and the emergence of synchronisation in multilayer neuronal models \cite{majhi2017chimera}.

A more unsual application of the hidden network concept is for decomposing complex single layer networks into simpler two-layer structures. One such example is second-neighbour preferential attachment; an implementation of the Barab\'{a}si-Albert model where nodes attach proportionally to the number of nodes within two steps of a target node \cite{falkenberg2020identifying}. Using our framework, the model is decomposed into an observed network, and a hidden network (in this case referred to as the influence network) where nodes are connected to all nodes which are two or fewer steps away, representing the node's sphere of influence. Here, second-neighbour preferential attachment is equivalent to conventional first-neighbour preferential attachment followed by a local copying step. Structural heterogeneity that is intrinsic in such a model has profound consequences for the time dependence of network growth \cite{falkenberg2020identifying}. 

\section{Conclusion}

We have introduced a general model of heterogeneous copying, implemented using a hidden network model. In the case of extreme copying bias, we have derived analytical results and have demonstrated significant differences to similar models with uniform copying rules. In particular, power-law degree distributions observed in uniform copying can be suppressed under heterogeneous copying, and networks are significantly more clustered if copying is biased towards a node's inner circle. Although a systematic study of copying in real networks is necessary, evidence suggests that heterogeneous copying may be relevant in a social context.

The heterogeneous copying model is just one simple application of a hidden network model.  In general, the framework allows us to deconstruct network growth heterogeneities in a non-arbitrary way, focusing on structural rather than node heterogeneity, and poses questions concerning the role of hidden information in network growth. Exploring these questions is a key aim in upcoming work.

\section{Code availability.} Python code is available at: \\ \href{https://github.com/MaxFalkenberg/RandomCopying}{github.com/MaxFalkenberg/RandomCopying}.

\section{Data availability.} All data can be generated using the Python code provided.

\section{Acknowledgements.} I am grateful to Tim S. Evans, Chester Tan and Kim Christensen for a number of useful discussions, and to Tim S. Evans and Chester Tan for proof reading the manuscript. I acknowledge a Ph.D. studentship from the Engineering and Physical Sciences Research Council through Grant No. EP/N509486/1.

\bibliography{apssamp}

\end{document}